\documentclass{vldb}

\input epsf

\begin{document}

\title{VOODB: A Generic Discrete-Event Random Simulation Model to Evaluate the Performances of OODBs}

\author{J\'{e}r\^{o}me Darmont~$^{\mbox{\tiny{$\dagger$}}}$~~~~~
        Michel Schneider~$^{\mbox{\tiny{$\ddagger$}}}$
}

\institution{Laboratoire d'Informatique (LIMOS)\\
             Universit\'{e} Blaise Pascal -- Clermont-Ferrand~II\\
             Complexe Scientifique des C\'{e}zeaux\\
             63177 Aubi\`{e}re Cedex\\
             FRANCE\\
             {\em $^{\mbox{\tiny{$\dagger$}}}$~darmont@libd2.univ-bpclermont.fr~~~~~
             $^{\mbox{\tiny{$\ddagger$}}}$~schneider@cicsun.univ-bpclermont.fr}\\}

\pagestyle{empty}
             
\maketitle

\begin{abstract}

Performance of object-oriented database systems (OODBs) is still an issue to both designers and users nowadays. The aim of this paper is to propose a generic discrete-event random simulation model, called VOODB, in order to evaluate the performances of OODBs in general, and the performances of optimization methods like clustering in particular. Such optimization methods undoubtedly improve the performances of OODBs. Yet, they also always induce some kind of overhead for the system. Therefore, it is important to evaluate their exact impact on the overall performances. VOODB has been designed as a generic discrete-event random simulation model by putting to use a modelling approach, and has been validated by simulating the behavior of the O$_{\mbox{\tiny{2}}}$ OODB and the Texas persistent object store. Since our final objective is to compare object clustering algorithms, some experiments have also been conducted on the DSTC clustering technique, which is implemented in Texas. To validate VOODB, performance results obtained by simulation for a given experiment have been compared to the results obtained by benchmarking the real systems in the same conditions. Benchmarking and simulation performance evaluations have been observed to be consistent, so it appears that simulation can be a reliable approach to evaluate the performances of OODBs.


\end{abstract}

\section{Introduction}

The needs in terms of performance evaluation for Object-Oriented Database Management Systems (OODBMSs) remain strong for both designers and users. Furthermore, it appears a necessity to perform {\em a priori} evaluations (before a system is actually built or achieved) in a variety of situations. A system designer may need to {\em a priori} test the efficiency of an optimization procedure or adjust the parameters of a buffering technique. It is also very helpful to users to {\em a priori} estimate whether a given system is able to handle a given workload.

The challenge of comparing object clustering techniques motivated us to contribute to OODBMSs performance evaluation. The principle of clustering is to store related objects close together on secondary storage. Hence, when one of these objects is loaded into the main memory, all its related objects are also loaded at the same time. Subsequent accesses to these objects are thus main memory accesses that are much faster than disk I/Os. However, clustering induces an overhead for the system (e.g., to reorganize the database, to collect and maintain usage statistics...), so it is important to gauge its true impact on the overall performances. For this particular problem, {\em a priori} evaluation is very attractive since it avoids coding inefficient algorithms in existing systems.

Discrete-event random simulation constitutes a traditional approach to {\em a priori} performance evaluation. Numerous simulation languages and/or environments exist nowadays. They allow the simulation of various classes of systems (computer systems, networks, production systems...). However, the use of simulation is not as widely disseminated as it could be in the database domain. The main difficulty is to elaborate a "good" functioning model for a system. Such a good model must be representative of the performances to evaluate, with the requested precision degree. For this sake, finding out the significant characteristics of a system and translating them into entities in the chosen simulation language often remains a specialist issue. Hence, users must call on consulting or specialized firms, which stretches out study times and costs.

In the field of OODBs, discrete-event random simulation has been chiefly used to validate proposals 
concerning optimization techniques, especially object clustering techniques. For instance, a dedicated model in PAWS was proposed in \cite{CHAN89} to validate a clustering and a buffering strategy in a CAD context. The objective was to find out how different optimization algorithms influence performances when the characteristics of the application accessing data vary, and which relationship exists between object clustering and parameters such as read/write ratio. Discrete-event random simulation was also used by \cite{DARM96, GAY97} in order to compare the efficiency of different clustering strategies for OODBs. The proposed models were coded in SLAM II.

Some other studies use simulation approaches that are not discrete-event random simulation approaches, but 
are nevertheless interesting. \cite{CHEN91} conducted simulation to show the effectiveness of different 
clustering schemes when parameters such as read/write ratio vary. The authors particularly focused on disk 
drive modelling. The CLAB ({\em CLustering LAboratory}) software \cite{TSAN92} was designed to compare graph partitioning algorithms applied to object clustering. It is constituted of a set of Unix tools programmed in C++, which can be assembled in various configurations. Yet other studies from the fields of distributed or parallel databases prove helpful, e.g., the modelling methodologies from \cite{IAEZ95} or the workload models from \cite{HE93, BATE95}.

These different studies bring forth the following observations.

First, most proposed simulation models are dedicated: they have been designed to evaluate the performance of a given optimization method. Furthermore, they only exploit one type of OODBMS, while various architectures influencing performances are possible (object server, page server, etc.). We advocate a more generic approach that would help modelling the behavior of various systems, implanting various object bases into these systems, and executing various transactions on these databases.

Besides, precision in specifications for these simulation models varies widely. It is thus not always easy to reproduce these models from the published material. Hence, it appears beneficial to make use of a modelling methodology that allows, step by step, analyzing a system and specifying a formalized knowledge model that can be distributed and reused.

Finally, as far as we know, none of these models has been validated. The behavior of the studied algorithm, if it is implemented in a real system, is thus not guaranteed to be the same than in simulation, especially concerning performance results. Confronting simulated results to measurements performed in the same conditions on a real system is a good method to hint whether a simulation model actually behaves like the system it models or not.

Considering these observations, our motivation is to propose a discrete-event random simulation model that 
addresses the issues of genericity, reusability and reliability. This model, baptized VOODB 
({\em Virtual Object-Oriented Database}), is indeed able to take into account different kinds of Client-Server architectures. It can also be parameterized to serve various purposes, e.g., to evaluate how a system reacts to different workloads or to evaluate the efficiency of optimization methods. Eventually, VOODB has been validated by confronting simulation results to performance measures achieved on real systems (namely O$_{\mbox{\tiny{2}}}$ and Texas).

The remainder of this paper is organized as follows. Section~2 introduces our modelling approach. Section~3
details the VOODB simulation model. Section~4 presents validation experiments for this model. We eventually 
conclude this paper and provide future research directions in Section~5.

\section{Modelling approach}

\begin{figure}[ht]
\begin{center}
\epsfxsize=5.5cm
\centerline{\epsffile{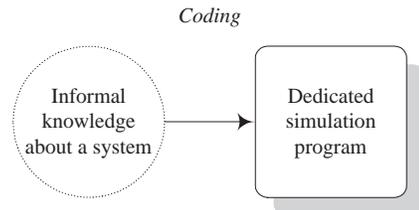}}
\caption{{\small Unstructured approach to simulation}}
\label{fig-1}
\end{center}
\end{figure}

In order to clearly identify the interest of a structured approach, let us imagine that a simulation program
is directly elaborated from informal knowledge concerning the studied system (Figure~\ref{fig-1}). Only experts mastering both the system to model and the target simulation language can satisfactorily use such an approach. It is thus only usable for punctual studies on relatively simple systems. The obtained simulation program is not meant to be reusable or later modified, and its documentation is minimal at best.

\begin{figure}[ht]
\begin{center}
\epsfxsize=8cm
\centerline{\epsffile{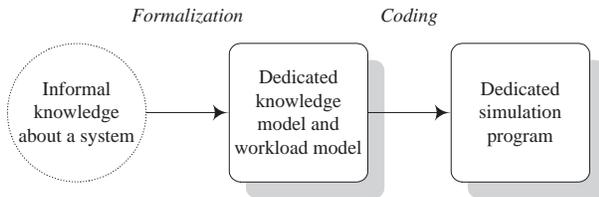}}
\caption{{\small Structured modelling approach}}
\label{fig-2}
\end{center}
\end{figure}

In opposition, a structured approach first consists in translating informal knowledge into an organized
knowledge model (Figure~\ref{fig-2}). This knowledge model rests on concepts close to those of the study domain. It may be more or less formalized, and must enable the systematic generation of a simulation program. This approach helps focusing on the modelled system's properties and to make abstractions of constraints related to the simulation environment. It facilitates feedback to improve simulation quality: it is possible to reconsider functioning hypothesis or detail some pieces by modifying the knowledge model and generating new code. Low-level parameters may be introduced (e.g., mean access time to a disk block). The workload model may be directly included into the knowledge model and may itself incorporate some parameters (e.g., the proportion of objects accessed within a given class). Since long, specialists in simulation worked on defining the principles of such an approach \cite{SARG79, NANC81, SARG91, BALC92, GOUR92, KELL97}.

\begin{figure*}[ht]
\begin{center}
\epsfxsize=11cm
\centerline{\epsffile{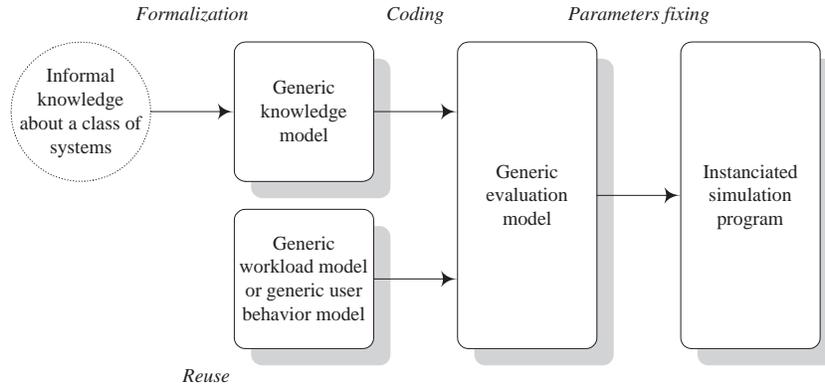}}
\caption{{\small Generic, structured modelling approach}}
\label{fig-3}
\end{center}
\end{figure*}

The approach we recommend (Figure~\ref{fig-3}) is a generic extension to the former approach. Its consists 
in broadening the study field to take into account a whole class of systems. The knowledge model must hence 
be tunable (e.g., high-level parameters may help selecting the system's architecture) and modular (some 
functionalities are included in specific modules that may be added or removed at will). The knowledge model, 
which is necessarily more complex, must be described in a hierarchical way up to the desired detail level.
We used the concepts and diagrams of UML \cite{RATI97} to describe it.

We also propose that the workload model be separately characterized. It is then possible to reuse workload 
models from existing benchmarks (like HyperModel \cite{ANDE90}, OO1 \cite{CATT91} or OO7 \cite{CARE93}) or 
establish a specific model. We chose to incorporate the workload model from the OCB 
({\em Object Clustering Benchmark}) generic benchmark \cite{DARM98}. Thanks to numerous parameters, this workload model can be adapted to various situations (existing benchmark workload, specific application workload...).

The generic simulation program is obtained in a systematic way. Its modular architecture is the result of the two models it is based on. The final simulation program for a specific case study is obtained by instantiation of this generic program. This approach guarantees a good reusability. It is possible after a first simulation experiment to broaden the study specter by changing the parameters' values (especially those concerning the workload), by selecting other modules (for instance, by replacing a clustering module by another), or by incorporating new modules.

\section{The VOODB simulation model}

\subsection{Knowledge model}

\begin{figure*}[ht]
\begin{center}
\epsfxsize=12.5cm
\centerline{\epsffile{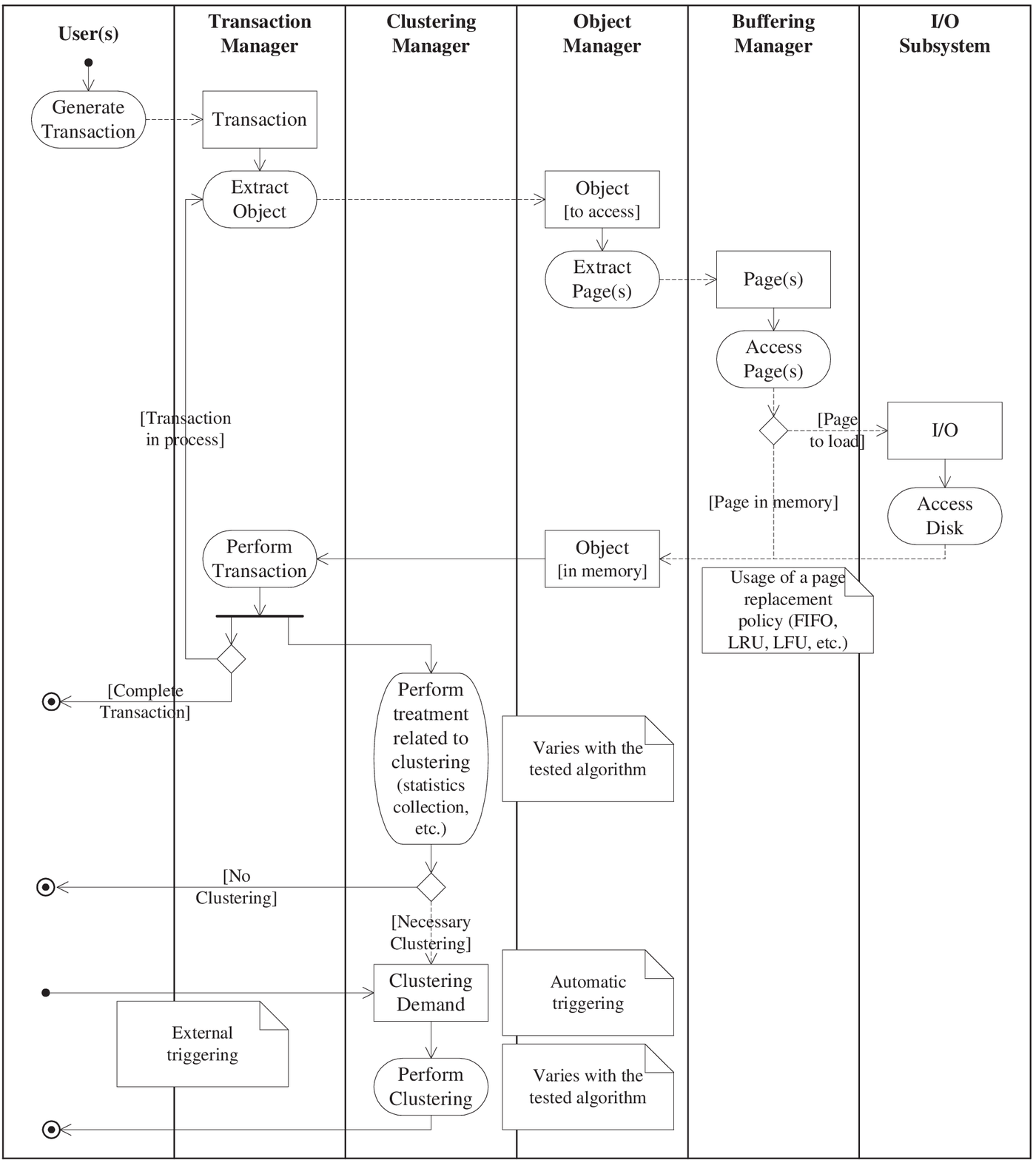}}
\caption{{\small Knowledge model}}
\label{fig-4}
\end{center}
\end{figure*}

In our context, the knowledge model describes the execution of transactions in an OODBMS (Figure~\ref{fig-4}).

Transactions are generated by the {\em Users}, who submit them to the {\em Transaction Manager}. The 
{\em Transaction Manager} determines which objects need to be accessed for the current transaction, and 
performs the necessary operations on these objects. A given object is requested by the {\em Transaction Manager}
to the {\em Object Manager} that finds out which disk page contains the object. Then, it requests the page 
from the {\em Buffering Manager} that checks if the page is present in the memory buffer. If not, it 
requests the page from the {\em I/O Subsystem} that deals with physical disk accesses. After an operation on
a given object is over, the {\em Clustering Manager} may update some usage statistics for the database. An 
analysis of these statistics can trigger a reclustering, which is then performed by the {\em Clustering Manager}.
Such a database reorganization can also be demanded externally by the {\em Users}. The only treatments that 
differ when two distinct clustering algorithms are tested are those performed by the {\em Clustering Manager}. 
Other treatments in the model remain the same, whether clustering is used or not, and whatever the 
clustering strategy.

\begin{figure}[ht]
\begin{center}
\epsfxsize=5cm
\centerline{\epsffile{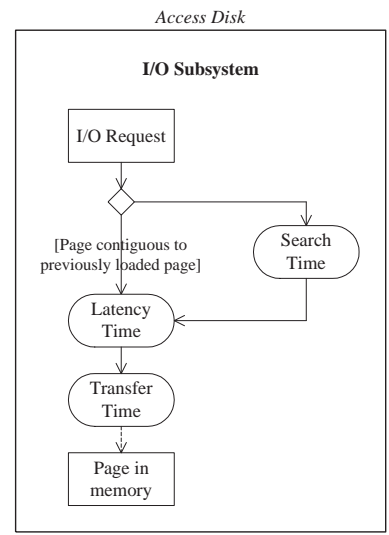}}
\caption{{\small "Access disk" functioning rule detail}}
\label{fig-5}
\end{center}
\end{figure}

The knowledge model is hierarchical. Each of its activities (rounded boxes) can be further detailed, as is illustrated in Figure~\ref{fig-5} for the "Access Disk" functioning rule.

The system's physical resources that appear as {\em swimlanes} in the knowledge model may be qualified as 
{\em active resources} since they actually perform some task. However, the system also includes 
{\em passive resources} that do not directly perform any task, but are used by the active resources to perform theirs. These passive resources do not appear on Figure~\ref{fig-4}, but must nevertheless be exhaustively listed (Table~\ref{tab-1}).

\begin{table}[ht]
\begin{center}
\bigskip
\begin{tabular}{|p{8cm}|}
\hline
{\small {\bf Passive resource}}\\ \hline\hline
{\small {\em Processor and main memory} in a centralized architecture, or {\em server processor and main memory} in a Client-Server architecture} \\ \hline
{\small {\em Clients processor and main memory} in a Client-Server architecture} \\ \hline
{\small {\em Server disk controller and secondary storage}} \\ \hline
{\small {\em Database}. Its concurrent access is managed by a scheduler that applies a transaction scheduling policy that depends on the multiprogramming level.} \\ \hline
\end{tabular}
\caption{{\small VOODB passive resources}}
\label{tab-1}
\end{center}
\end{table}

\subsection{Evaluation model}

\subsubsection{Simulator selection}

We first selected the QNAP2 (Queuing Network Analysis Package 2$^{\mbox{\tiny{nd}}}$ generation, version 9) discrete-event random simulation software \cite{SIMU95} to implement VOODB, because it proposes the following essential features:
\begin{itemize}
\item QNAP2 is a validated and reliable simulation tool;
\item QNAP2 allows the use of an object-oriented approach (since version 6);
\item QNAP2 includes a full algorithmic language, derived from Pascal, which allows a relatively easy implementation of complex algorithms (object clustering, buffer page replacement, prefetching, etc.).
\end{itemize}

However, QNAP2 is an interpreted language. The models written in QNAP2 are hence much slower at execution time than if they were written in a compiled language. Therefore, we could not achieve the intensive simulation campaign we intended to. For instance, the simplest simulation experiments (without clustering) were 8 hours long, while the most complex were more than one week long. Thus, we could not gain much insight beyond basic results.

We eventually considered the use of C++, which is both an object-oriented and compiled language. This also 
allowed us reusing most of the OCB benchmark's C++ code. But the existing C++ simulation packages were 
either not completely validated, featured much more than we actually needed, and hence were getting as 
complicated to use as general simulation languages, or were not free. Hence, we decided to design our own 
C++ simulation kernel. It has been baptized DESP-C++ ({\em Discrete-Event Simulation Package for C++}). Its main characteristics are validity, simplicity and efficiency. DESP-C++ has been validated by comparing the results of several simulation experiments conducted with DESP-C++ and QNAP2. Simulation experiments are now 20 to 1,000 times quicker with DESP-C++, depending on the model's complexity (the more a model is complex, the more QNAP2 performs poorly).

\subsubsection{Knowledge model translation}

Once the knowledge model is designed, it can be quasi-automatically translated into an 
evaluation model using any environment, whether it is a general simulation language or a usual 
programming language. Each entity in the knowledge model appears in the evaluation model some 
way. In an object-oriented environment, resources (active and passive) become instantiated 
classes, and functioning rules are translated into methods.

More precisely, the translation from the knowledge model to the evaluation model proceeds as follows:
\begin{itemize}
\item each active resource ({\em swimlanes} in Figure~\ref{fig-4}) becomes a component of the simulation program (i.e., a class);
\item each object (square boxes in Figure~\ref{fig-4}) becomes an interface to these components (i.e., it is used as a parameter in messages between two classes);
\item each activity (round boxes in Figure~\ref{fig-4}) becomes a method within a component.
\end{itemize}
	
Passive resources are classes bearing mainly two methods: one to reserve the resource and another one to release it.

Table~\ref{tab-2} recapitulates how entities from the knowledge model are translated in QNAP2 and DESP-C++, which both use a resource view (where the demeanor of each active resource is described). Table~\ref{tab-2} also provides a translation in SLAM II \cite{PRIT86}, which uses a transaction view (where the specification concerns the operations undergone by the entities flowing through the system). This is simply to show that the implementation of VOODB with a simulator using the transaction view is also possible.

\begin{table*}[ht]
\begin{center}
\bigskip
\begin{tabular}{|l|l|p{3.5cm}|p{3.5cm}|p{3.5cm}|}
\hline
{\small {\bf Subsystem}} &
{\small {\bf Entity}} &
{\small {\bf QNAP2 translation}} &
{\small {\bf DESP-C++ translat.}} &
{\small {\bf SLAM II translation}} \\ \hline \hline
{\small Workload} &
{\small (Sub)Transaction} &
{\small CUSTOMER object} &
{\small Instance of class Client} &
{\small SLAM Entity} \\ \hline
{\small Physical} &
{\small Passive resource} &
{\small RESOURCE STATION object} &
{\small Instance of class Resource} &
{\small RESOURCE block} \\ \cline{2-5}
{\small ~}&
{\small Active resource} &
{\small Standard STATION object} &
{\small Instance of an active resource class inheriting from class Resource} &
{\small Set of SLAM nodes (ACTIVITY, EVENT, FREE, GOON...)} \\ \hline
{\small Control} &
{\small Functioning rule} &
{\small PROCEDURE called in the SERVICE clause of an active resource} &
{\small Method of an active resource class} &
{\small FORTRAN subroutine called in an EVENT node} \\ \hline
\end{tabular}
\caption{{\small Translation of the knowledge model entities}}
\label{tab-2}
\end{center}
\end{table*}

\subsection{Genericity in VOODB}

Genericity in VOODB is primarily achieved through a set of parameters that help tuning the model in a variety of configurations, and setting up the different policies influencing the eventual behavior of an instance of the generic evaluation model. VOODB also benefits from the genericity of the OCB benchmark \cite{DARM98} at the workload level, since OCB is itself tunable through a thorough set of 26 parameters. The parameters defining an instance of the VOODB evaluation model are presented in Table~\ref{tab-3}. Each active resource is actually associated to a set of parameters. These parameters are normally directly deduced from the studied system's specifications. However, some parameters are not always readily available and have to be worked out from benchmarks or measures (e.g., to determine network throughput or disk performances).

\begin{table*}[ht]
\begin{center}
\bigskip
\begin{tabular}{|l|p{3.75cm}|l|p{4cm}|p{2cm}|}
\hline
{\small {\bf Active resource}} &
{\small {\bf Parameter}} &
{\small {\bf Code}} &
{\small {\bf Range}} &
{\small {\bf Default}} \\ \hline \hline
{\small System} &
{\small System class} &
{\small SYSCLASS} &
{\small \{Centralized $|$ Object Server $|$ Page Server $|$ DB Server $|$ Other\}} &
{\small Page Server} \\ \cline{2-5}
{\small ~} &
{\small Network throughput} &
{\small NETTHRU} &
{\small --} &
{\small 1 MB/s} \\ \hline
{\small Buffering Manager} &
{\small Disk page size} &
{\small PGSIZE} &
{\small \{512 $|$ 1024 $|$ 2048 $|$ 4096 \} bytes} &
{\small 4096 bytes} \\ \cline{2-5}
{\small ~} &
{\small Buffer size} &
{\small BUFFSIZE} &
{\small --} &
{\small 500 pages} \\ \cline{2-5}
{\small ~} &
{\small Buffer page replacement strategy} &
{\small PGREP} &
{\small \{RANDOM $|$ FIFO $|$ LFU $|$ LRU-K $|$ CLOCK $|$ GCLOCK $|$ Other\}} &
{\small LRU-1} \\ \cline{2-5}
{\small ~} &
{\small Prefetching policy} &
{\small PREFETCH} &
{\small \{None $|$ Other\}} &
{\small None} \\ \hline
{\small Clustering Manager} &
{\small Object clustering policy} &
{\small CLUSTP} &
{\small \{None $|$ Other\}} &
{\small None} \\ \cline{2-5}
{\small ~} &
{\small Objects initial placement} &
{\small INITPL} &
{\small \{Sequential $|$ Optimized sequential $|$ Other\}} &
{\small Optimized Sequential} \\ \hline
{\small I/O Subsystem} &
{\small Disk search time} &
{\small DISKSEA} &
{\small --} &
{\small 7.4 ms} \\ \cline{2-5}
{\small ~} &
{\small Disk latency time} &
{\small DISKLAT} &
{\small --} &
{\small 4.3 ms} \\ \cline{2-5}
{\small ~} &
{\small Disk transfer time} &
{\small DISKTRA} &
{\small --} &
{\small 0.5 ms} \\ \hline
{\small Transaction Manager} &
{\small Multiprogramming level} &
{\small MULTILVL} &
{\small --} &
{\small 10} \\ \cline{2-5}
{\small ~} &
{\small Locks acquisition time} &
{\small GETLOCK} &
{\small --} &
{\small 0.5 ms} \\ \cline{2-5}
{\small ~} &
{\small Locks release time} &
{\small RELLOCK} &
{\small --} &
{\small 0.5 ms} \\ \hline
{\small Users} &
{\small Number of users} &
{\small NUSERS} &
{\small --} &
{\small 1} \\ \hline
\end{tabular}
\caption{{\small VOODB parameters}}
\label{tab-3}
\end{center}
\end{table*}

Our generic model allows simulating the behavior of different types of OODBMSs. It is in 
particular adapted to the different configurations of Client-Server architectures, which are 
nowadays the standard in OODBs. Our model is actually especially suitable to page server 
systems (like ObjectStore \cite{LAMB91}, or O$_{\mbox{\tiny{2}}}$ \cite{DEUX91}), but can also 
be used to model object server systems (like ORION \cite{KIM88} or ONTOS \cite{ANDR91}), or 
database server systems, or even multiserver hybrid systems (like GemStone \cite{SERV92}). The 
organization of the VOODB components is controlled by the "System class" parameter.

\section{Validation experiments}

\subsection{Experiments scope}

Though we use validated tools (QNAP2, or DESP-C++), the results provided by simulation are not guaranteed to be consistent with reality. To check out if our simulation models were indeed valid, we simulated the behavior of two systems that offer object persistence: O$_{\mbox{\tiny{2}}}$ \cite{DEUX91} and Texas \cite{SING92}. We compared these results to those provided by benchmarking these real systems with OCB. The objective here was to use the same workload model in both sets of experiments.

In a second step, we seeked to evaluate the impact of an optimization method (the DSTC clustering technique \cite{BULL96}, which has been implemented in Texas). We again compared results obtained by simulation and direct measures performed under the same conditions on the real system.

Due to space constraints, we only present here our most significant results. Besides, our goal is not to perform sound performance evaluations of O$_{\mbox{\tiny{2}}}$, Texas and DSTC. We just seek to show our simulation approach can provide trustworthy results.

\subsection{Experimental conditions}

\subsubsection{Real systems}

The O$_{\mbox{\tiny{2}}}$ server we used (version~5.0) is installed on an IBM RISC~6000~43P240 biprocessor 
workstation. Each processor is a Power~PC 604e~166. The workstation has 1~GB ECC RAM. Its operating system is 
AIX version~4. The O$_{\mbox{\tiny{2}}}$ server cache size is 16~MB by default.

The version of Texas we use is a prototype (version~0.5) running on a PC~Pentium-II~266 with 64~MB of SDRAM, 
which operating system is Linux, version~2.0.30. The swap partition size is 64~MB. DSTC is integrated in 
Texas as a collection of new modules, and a modification of several Texas modules. Texas and the additional 
DSTC modules were compiled using the GNU C++ (version~2.7.2.1) compiler.

\subsubsection{Simulation}

Our C++ simulation models were compiled with the GNU C++ (version~2.7.2.1) compiler. They run on a PC~Pentium-II~266
with 64~MB of SDRAM, under Windows~95.


In order to simulate the behavior of O$_{\mbox{\tiny{2}}}$ and Texas, VOODB has been parameterized as showed in Table~\ref{tab-7}.
These parameters were all fixed up from the specification and configuration of the hardware and software systems we used.

\begin{table*}[ht]
\begin{center}
\bigskip
\begin{tabular}{|l|l|c|c|}
\hline
{\small {\bf Parameter}} &
{\small {\bf Code}} &
{\small {\bf Value for O$_{\mbox{\tiny{2}}}$}} &
{\small {\bf Value for Texas}} \\ \hline \hline
{\small System class} &
{\small SYSCLASS} &
{\small Page server} &
{\small Centralized} \\ \hline
{\small Network throughput} &
{\small NETTHRU} &
{\small +$\infty$} &
{\small N/A} \\ \hline
{\small Disk page size} &
{\small PGSIZE} &
{\small 4096 bytes} &
{\small 4096 bytes} \\ \hline
{\small Buffer size} &
{\small BUFFSIZE} &
{\small 3840 pages} &
{\small 3275 pages} \\ \hline
{\small Buffer page replacement strategy} &
{\small PGREP} &
{\small LRU} &
{\small LRU} \\ \hline
{\small Prefetching policy} &
{\small PREFETCH} &
{\small None} &
{\small None} \\ \hline
{\small Object clustering policy} &
{\small CLUSTP} &
{\small None} &
{\small DSTC} \\ \hline
{\small Objects initial placement} &
{\small INITPL} &
{\small Optimized Sequential} &
{\small Optimized Sequential} \\ \hline
{\small Disk search time} &
{\small DISKSEA} &
{\small 6.3 ms} &
{\small 7.4 ms}  \\ \hline
{\small Disk latency time} &
{\small DISKLAT} &
{\small 2.99 ms} &
{\small 4.3 ms} \\ \hline
{\small Disk transfer time} &
{\small DISKTRA} &
{\small 0.7 ms} &
{\small 0.5 ms} \\ \hline
{\small Multiprogramming level} &
{\small MULTILVL} &
{\small 10} &
{\small 1} \\ \hline
{\small Locks acquisition time} &
{\small GETLOCK} &
{\small 0.5 ms} &
{\small 0} \\ \hline
{\small Locks release time} &
{\small RELLOCK} &
{\small 0.5 ms} &
{\small 0} \\ \hline
{\small Number of users} &
{\small NUSERS} &
{\small 1} &
{\small 1} \\ \hline
\end{tabular}
\caption{{\small Parameters defining the O$_{\mbox{\tiny{2}}}$ and the Texas systems within VOODB}}
\label{tab-7}
\end{center}
\end{table*}


Our simulation results have been achieved with 95\% confidence intervals ($c=0.95$). To determine
these intervals, we used the method exposed in \cite{BANK96}. For given observations,
sample mean $\bar{X}$ and sample standard deviation $\sigma$ are computed. The half-interval width
$h$ is $h$=$t$$_{\mbox{\tiny{n-1,1-$\alpha$/2}}}$.$\sigma$/$\sqrt{n}$,
where $t$ is given by the Student
$t$-distribution, $n$ is the number of replications and $\alpha$=$1-c$. The mean value belongs 
to the [$\bar{X}$-$h$,$\bar{X}$+$h$] confidence interval with a probability $c=0.95$.

Since we wish to be within 5\% of the sample mean
with 95\% confidence, we first performed a pilot study with $n=10$. Then we computed the number
of necessary additional replications $n$$^{\mbox{\tiny{*}}}$ using the equation: 
$n$$^{\mbox{\tiny{*}}}$=$n$.($h$/$h$$^{\mbox{\tiny{*}}}$)$^{\mbox{\tiny{2}}}$, where $h$ is the
half-width of the confidence interval for the pilot study and $h$$^{\mbox{\tiny{*}}}$ the half-width
of the confidence interval for all replications (the desired half-width).
  
Our simulation results showed that the required precision was achieved for all our performance
criteria when $n$+$n$$^{\mbox{\tiny{*}}}$$\geq$100, with a broad security margin. We thus performed 100 replications in
all our experiments. In order to preserve results clarity in the following figures, we did
not include the confidence intervals. They are however computed by default by DESP-C++.
  
\subsection{Experiments on O$_{\mbox{\tiny{2}}}$ and Texas}

First, we investigated the effects of the object base size (number of classes and number of instances in the 
database) on the performances (mean number of I/Os necessary to perform the transactions) of the studied systems. 
In this series of experiments, the number of classes  in the schema ({\em NC}) is 20 or 50, and the number 
of instances ({\em NO}) varies from 500 to 20,000. The workload configuration is showed in Table~\ref{tab-8}. The other OCB parameters were set up to their default values.

\begin{table*}[ht]
\begin{center}
\bigskip
\begin{tabular}{|l|c|l|c|}
\hline
{\small {\bf Parameter}} &
{\small {\bf Val.}} &
{\small {\bf Parameter}} &
{\small {\bf Val.}} \\ \hline \hline
{\small {\em COLDN:} Number of transactions (cold run)} &
{\small 0} &
{\small {\em HOTN:} Number of transactions (warm run)} &
{\small 1000} \\ \hline
{\small {\em PSET:} Set-oriented access occurrence probability} &
{\small 0.25} &
{\small {\em SETDEPTH:} Set-oriented access depth} &
{\small 3} \\ \hline
{\small {\em PSIMPLE:} Simple traversal occurrence probability} &
{\small 0.25} &
{\small {\em SIMDEPTH:} Simple traversal access depth} &
{\small 3} \\ \hline
{\small {\em PHIER:} Hierarchy traversal occurrence probability} &
{\small 0.25} &
{\small {\em HIEDEPTH:} Hierarchy traversal access depth} &
{\small 5}  \\ \hline
{\small {\em PSTOCH:} Stochastic traversal occurrence probability} &
{\small 0.25} &
{\small {\em STODEPTH:} Stochastic traversal access depth} &
{\small 50} \\ \hline
\end{tabular}
\caption{{\small OCB workload definition}}
\label{tab-8}
\end{center}
\end{table*}

In a second step, we varied the server cache size (O$_{\mbox{\tiny{2}}}$) or the available 
main memory (Texas) in order to study the effects on performances (mean 
number of I/Os). The objective was also to simulate the system's reaction when the (memory size
/ database size) ratio decreases. In the case of O$_{\mbox{\tiny{2}}}$, the server cache size 
is specified by environment variables. Our Texas version is implanted under Linux, which allows 
setting up memory size at boot time. Cache or main memory size varied from 8~MB to 64~MB in 
these experiments. Database size was fixed ({\em NC}=50, {\em NO}=20,000), we reused the workload from 
Table~\ref{tab-8}, and the other OCB parameters were set up to their default values.

\subsubsection{Results concerning O$_{\mbox{\tiny{2}}}$}

\paragraph{{\em Database size variation}}

\begin{figure}[ht]
\begin{center}
\epsfxsize=8cm
\centerline{\epsffile{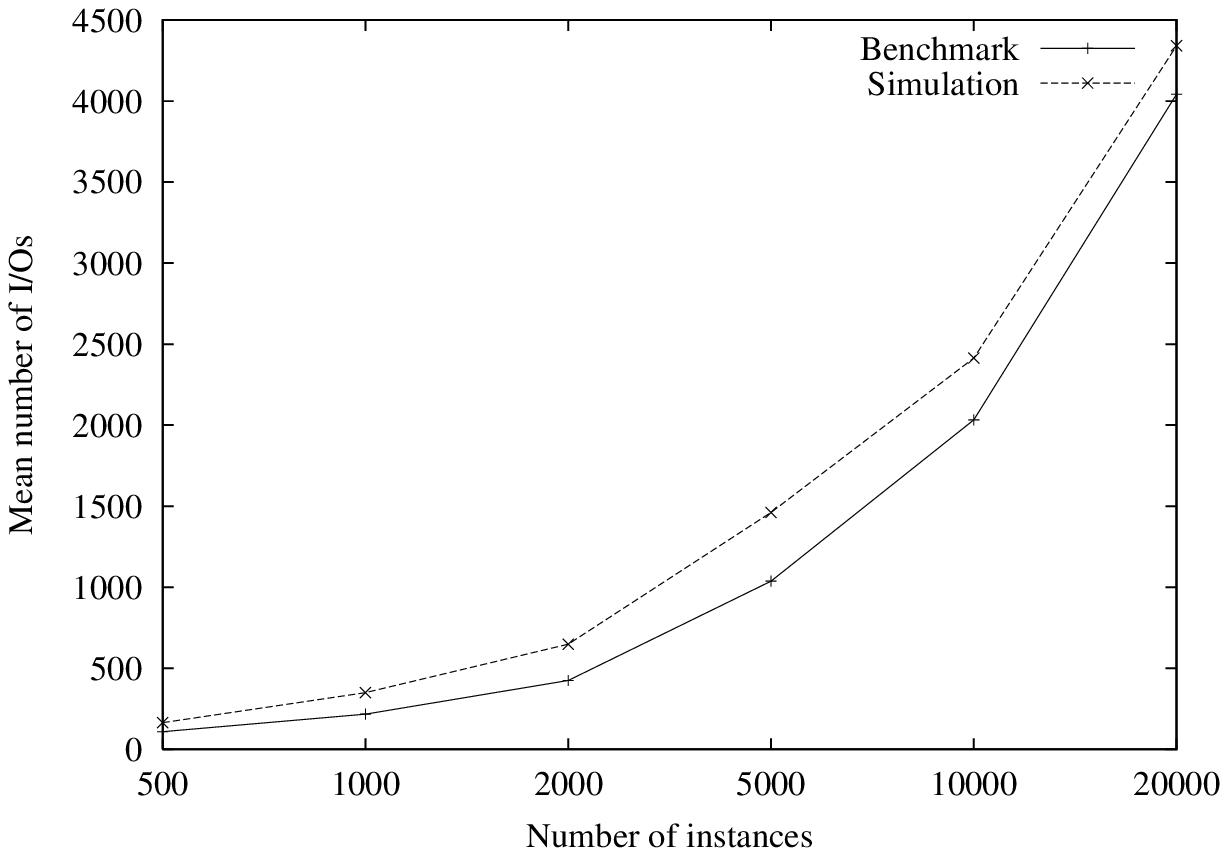}}
\caption{{\small Mean number of I/Os depending on number of instances (O$_{\mbox{\tiny{2}}}$ -- 20 classes)}}
\label{fig-8}
\end{center}
\end{figure}

\begin{figure}[ht]
\begin{center}
\epsfxsize=8cm
\centerline{\epsffile{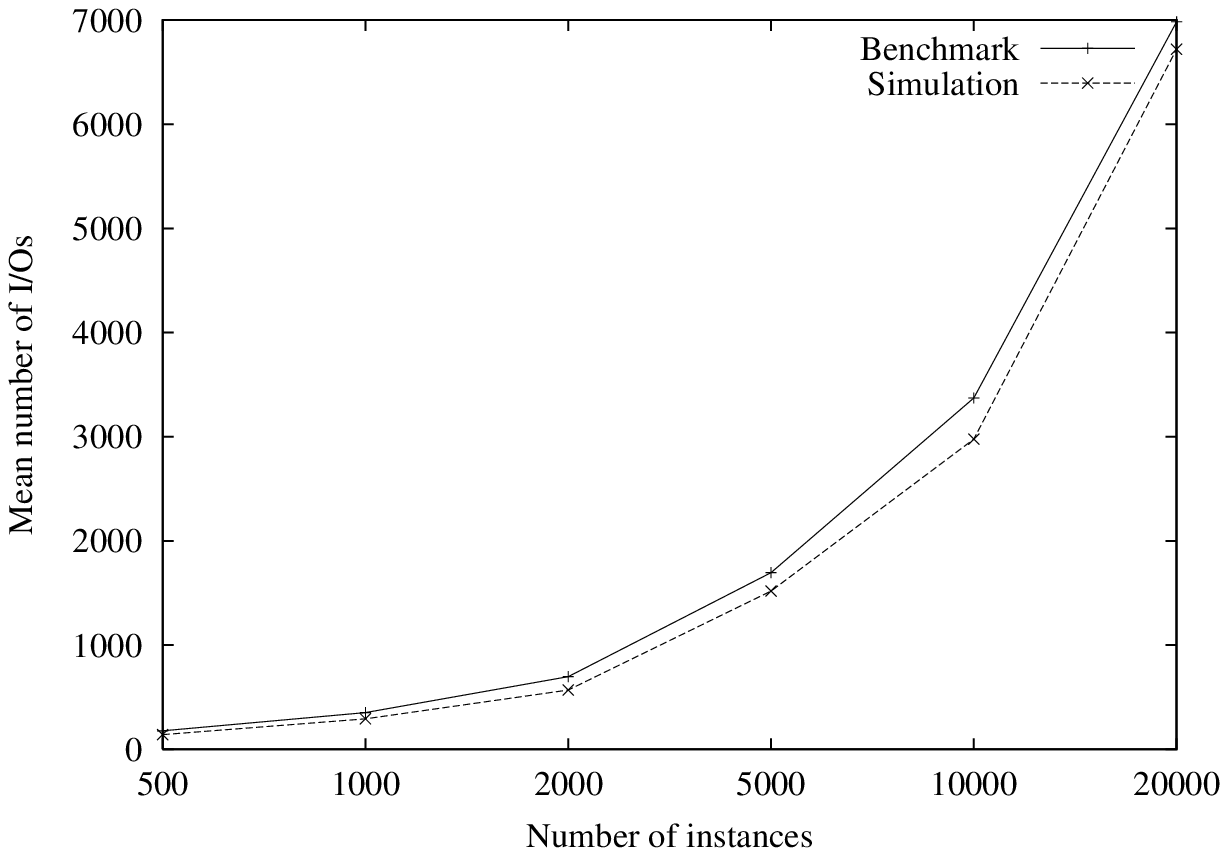}}
\caption{{\small Mean number of I/Os depending on number of instances (O$_{\mbox{\tiny{2}}}$ -- 50 classes)}}
\label{fig-9}
\end{center}
\end{figure}

Figures~\ref{fig-8} and \ref{fig-9} show how the performances of O$_{\mbox{\tiny{2}}}$ vary in terms of 
number of I/Os when the number of classes and the number of instances in the database vary. 
We can see that simulation results are in absolute value lightly different from the results measured on the
real system, but that they clearly show the same tendency. The behavior of VOODB is indeed conforming to reality.

\paragraph{{\em Cache size variation}}

\begin{figure}[ht]
\begin{center}
\epsfxsize=8cm
\centerline{\epsffile{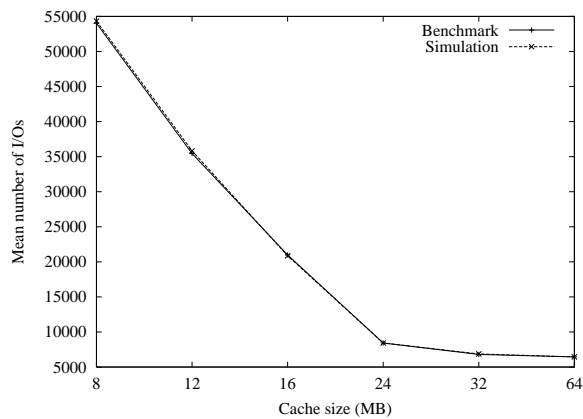}}
\caption{{\small Mean number of I/Os depending on cache size (O$_{\mbox{\tiny{2}}}$)}}
\label{fig-10}
\end{center}
\end{figure}

The results obtained in this experiment in terms of number of I/Os are presented in Figure~\ref{fig-10}. 
They show that the performances of O$_{\mbox{\tiny{2}}}$ rapidly degrade when the database size (about 28~MB 
on an average) becomes greater than the cache size. This decrease in performance is linear. 
Figure~\ref{fig-10} also shows that the performances of O$_{\mbox{\tiny{2}}}$ can be reproduced again with 
our simulation model.

\subsubsection{Results concerning Texas}

\paragraph{{\em Database size variation}}

\begin{figure}[ht]
\begin{center}
\epsfxsize=8cm
\centerline{\epsffile{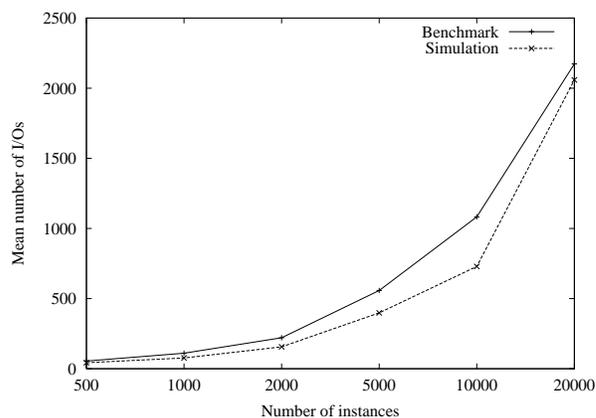}}
\caption{{\small Mean number of I/Os depending on number of instances (Texas -- 20 classes)}}
\label{fig-11}
\end{center}
\end{figure}

\begin{figure}[ht]
\begin{center}
\epsfxsize=8cm
\centerline{\epsffile{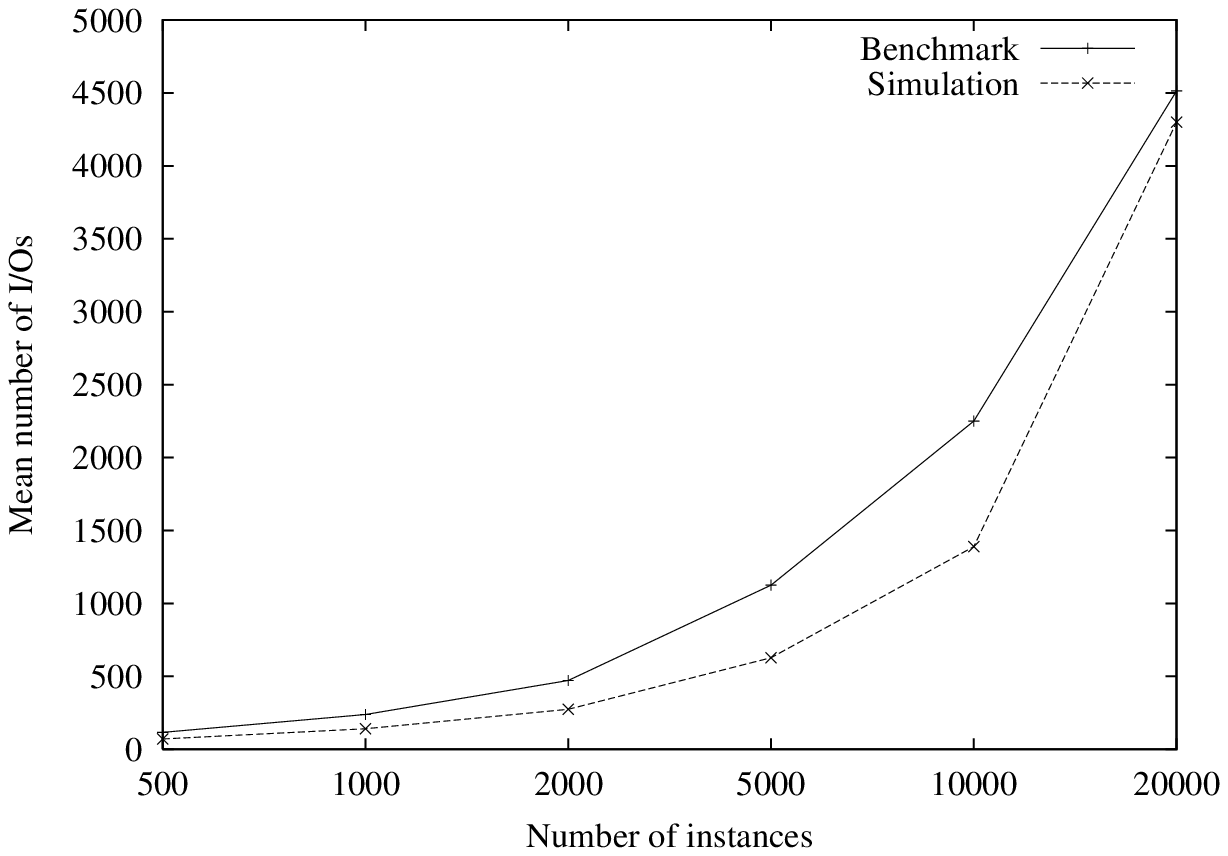}}
\caption{{\small Mean number of I/Os depending on number of instances (Texas -- 50 classes)}}
\label{fig-12}
\end{center}
\end{figure}

Figures~\ref{fig-11} and \ref{fig-12} show how the performances of Texas vary in terms of
number of I/Os when the number of classes and the number of instances in the database vary. As is the case 
with O$_{\mbox{\tiny{2}}}$, we can see that simulation results and results measured on the real system lightly differ in absolute value, but that they bear the same tendency.

\paragraph{{\em Memory size variation}}
	
\begin{figure}[ht]
\begin{center}
\epsfxsize=8cm
\centerline{\epsffile{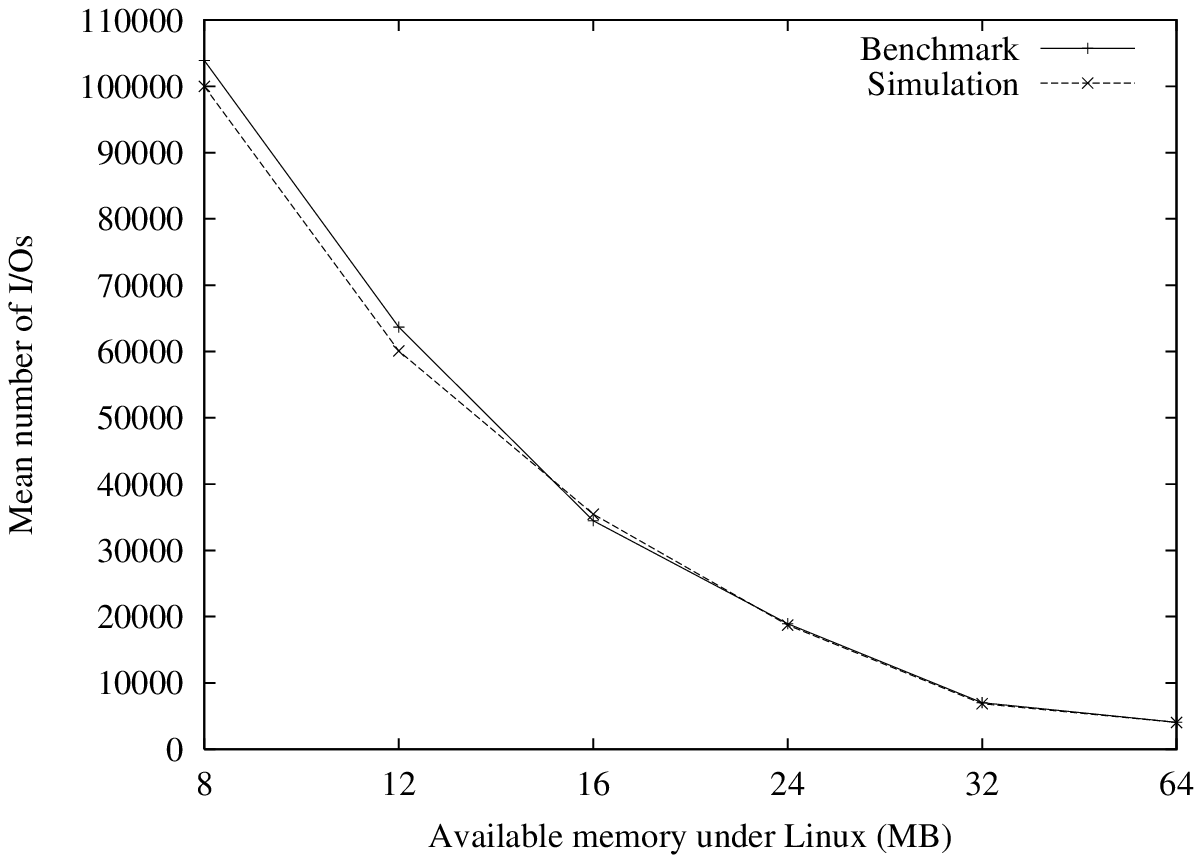}}
\caption{{\small Mean number of I/Os depending on memory size (Texas)}}
\label{fig-13}
\end{center}
\end{figure}

Since Texas uses the virtual memory mechanisms from the operating system, we studied the effects of a 
decrease in available main memory size under Linux. The results obtained in terms of number of I/Os 
are presented in Figure~\ref{fig-13}. They show that the performances of Texas rapidly degrade when the
main memory size becomes smaller than the database size (about 21~MB on an average). This degradation is 
due to Texas' object loading policy, which provokes the reservation in memory of numerous pages even before 
they are actually loaded. This process is clearly exponential and generates a costly swap, which is as 
important a hindrance as the main memory is small. The simulation results provided by VOODB are still 
conforming to reality.

\subsection{Effects of DSTC on the performances of Texas}

We underlined DSTC's clustering capability by placing the algorithm in favorable conditions. For this sake, we ran very characteristic transactions (namely, depth-3 hierarchy traversals) and measured the performances of Texas before and after clustering. We also evaluated clustering overhead. We checked out that the behavior of DSTC was the same in our simulation model and in the real system, by counting the number of created clusters and these clusters' mean size.

This experiment has been performed on a mid-sized database (50 classes, 20,000 instances, about 20~MB on an 
average). We had also planned to perform this experiment on a large object base, but we encountered 
technical problems with Texas/DSTC. To bypass the problems, we reduced the main memory size from 64~MB to 
8~MB so that the database size is actually large compared to the main memory size. Then, we reused the mid-sized object base from the first series of experiments. The other OCB parameters were set up to their default values.

Table~\ref{tab-9} 
presents the numbers of I/Os 
achieved on the real system and in simulation, for the mid-sized database. 
It shows that DSTC allows substantial performance improvements (performance gain around a factor 5). 
Clustering overhead is high, though. Furthermore, the simulation results are overall consistent with the 
performance measurements done on the real system, except concerning clustering overhead, which is far less 
important in simulation than in reality.

\begin{table}[ht]
\begin{center}
\bigskip
\begin{tabular}{|l|c|c|c|}
\hline
{\small {\bf ~}} &
{\small {\bf Bench.}} &
{\small {\bf Sim.}} &
{\small {\bf Ratio}} \\ \hline \hline
{\small Pre-clustering usage} &
{\small 1890.70} &
{\small 1878.80} &
{\small 1.0063} \\ \hline
{\small Clustering overhead} &
{\small 12799.60} &
{\small 354.50} &
{\small 36.1060} \\ \hline
{\small Post-clustering usage} &
{\small 330.60} &
{\small 350.50} &
{\small 0.9432} \\ \hline
{\small Gain} &
{\small 5.71} &
{\small 5.36} &
{\small 1.0652} \\ \hline
\end{tabular}
\caption{{\small Effects of DSTC on the performances (mean number of I/Os) -- Mid-sized base}}
\label{tab-9}
\end{center}
\end{table}


This flagrant inconsistency is not due to a bug in the simulation model, but to a particularity in Texas. 
Indeed, after reorganization of the database by DSTC, objects are moved on different disk pages. Hence, 
their OIDs change because Texas uses physical OIDs. In order to maintain consistency among inter-object references, the whole database must be scanned and all references toward moved objects must be updated. This phase, which is very costly both in terms of I/Os and time, is pointless in our simulation models, since they necessarily use logical OIDs.

To simulate DSTC's behavior within Texas in a wholly faithful way, it would have been easy to take this conversion time into account in our simulations. However, we preferred keeping our initial results in order to underline the difficulty to implant a dynamic clustering technique within a persistent object store using physical OIDs. On the other hand, our simulations show that such a dynamic technique is perfectly viable in a system with logical OIDs.

The number of clusters built by the DSTC method and these clusters' average size are presented in Table~\ref{tab-11}. We can observe again that there are few differences between the real system's behavior and its simulated behavior with VOODB.

\begin{table}[ht]
\begin{center}
\bigskip
\begin{tabular}{|l|c|c|c|}
\hline
{\small {\bf ~}} &
{\small {\bf Bench.}} &
{\small {\bf Sim.}} &
{\small {\bf Ratio}} \\ \hline \hline
{\small Mean number of clusters} &
{\small 82.23} &
{\small 84.01} &
{\small 0.9788} \\ \hline
{\small Mean number of obj./clust.} &
{\small 12.83} &
{\small 13.73} &
{\small 0.9344} \\ \hline
\end{tabular}
\caption{{\small DSTC clustering}}
\label{tab-11}
\end{center}
\end{table}

Eventually, Table~\ref{tab-12} 
presents the number of I/Os
achieved on the real system and by simulation, for the "large" database. 
It shows that simulation results are still consistent with performances observed on the real system. 
Furthermore, the gain induced by clustering is much higher when the database does not wholly fit into the 
main memory (increase from a factor 5 to a factor of about 30). This result was foreseeable, since the more 
the memory size is reduced, the more the system must perform page replacements. Unused pages hence normally 
remain only a short time in memory. A good object clustering is thus more useful in these conditions. 
Clustering overhead is not repeated here, since we reused the object base (in its initial and clustered 
state) from the first series of experiments.

\begin{table}[ht]
\begin{center}
\bigskip
\begin{tabular}{|l|c|c|c|}
\hline
{\small {\bf ~}} &
{\small {\bf Bench.}} &
{\small {\bf Sim.}} &
{\small {\bf Ratio}} \\ \hline \hline
{\small Pre-clustering usage} &
{\small 12504.60} &
{\small 12547.80} &
{\small 0.9965} \\ \hline
{\small Post-clustering usage} &
{\small 424.30} &
{\small 441.50} &
{\small 0.9610} \\ \hline
{\small Gain} &
{\small 29.47} &
{\small 28.42} &
{\small 1.0369} \\ \hline
\end{tabular}
\caption{{\small Effects of DSTC on the performances (mean number of I/Os) -- "Large" base}}
\label{tab-12}
\end{center}
\end{table}


\section{Conclusion}

We present in this paper a generic discrete-event random simulation model, VOODB, which is designed to 
evaluate the performances of OODBs. VOODB is parameterized and modular, and thus can be adapted to various purposes. It allows the simulation of various types of OODBMSs and can capture performance improvements achieved by optimization methods. Such optimization methods can be included in VOODB as interchangeable modules. Furthermore, the workload model adopted in VOODB (the OCB benchmark) can also be replaced by another existing benchmark or a specific workload. VOODB may be used as is (its C++ code is freely available) or tailored to fit some particular needs.

We have illustrated the genericity of VOODB and hinted its validity by setting its parameters to simulate the behavior of the O$_{\mbox{\tiny{2}}}$ OODB and the Texas persistent object store. We correlated the simulated performances of both systems with actual performance measures of the real systems (performed with the OCB benchmark), and observed they matched. The effects of the DSTC clustering technique on Texas' performances have also been mimicked by simulation.

VOODB may be used for several purposes. The performances of a single, or several optimization algorithms, may be evaluated in many different conditions. For instance, the host OODB or OS can vary, to see how a given algorithm behaves. Such clustering strategies may also be compared to each other that way. Furthermore, simulation has a low cost, since the different simulated systems (hardware, OS, OODBs) do not need to be acquired. Their specifications are enough. Eventually, we can {\em a priori} model the behavior of new systems, test their performances, analyze the simulation results, and ameliorate them (and then reiterate the process).

Eventually, VOODB has been obtained through the application of a modelling methodology that led to the design of a generic knowledge model and a generic evaluation model. This approach ensured that the specifications of the simulation models were precise enough for our deeds and that the evaluation model was properly translated from the knowledge model. It is also possible to reuse our knowledge model to produce simulation programs in other simulation languages or environment than QNAP2 or DESP-C++.

The reusability of VOODB may be important in a context of limited publicity. Since benchmarkers can encounter serious legal problems with OODB vendors if they publish performance studies \cite{CARE93}, it can be helpful to have a tool to perform private performance evaluations.

Future work concerning this study is first performing intensive simulation experiments with DSTC. We indeed only have basic results. It would be interesting to know the right value for DSTC's parameters in various conditions. We also plan to evaluate the performances of other optimization techniques, like the clustering strategy proposed by \cite{GAY97}, which has also been implemented in Texas, recently. This clustering technique originates from collaboration between the University of Oklahoma and Blaise Pascal University. The ultimate goal is to compare different clustering strategies, to determine which one performs best in a given set of conditions.

Though simulation may be used in substitution to benchmarking (mainly for {\em a priori} performance evaluations), it may also be used in complement to benchmarking. For instance, mixed benchmarking-simulation approach may be used to measure some performance criteria necessitating precision by experimentation, and other criteria by simulation (e.g., to determine the best architecture for a given purpose). With such an approach, using the same workload (e.g., OCB) in simulation and on the real system is essential.

The VOODB simulation model could also be improved, in order to include more components influencing the performances of OODBs. For instance, it currently only provides a few basic buffering strategies (RANDOM, FIFO, LFU, LRU-K, CLOCK...) and no prefetching strategy, which have been demonstrated to influence the performances of OODBs a lot, too \cite{BULL96}.

VOODB could even be extended to take into account completely different aspects of performance in OODBs, like concurrency control or query optimization. VOODB could also take into account random hazards, like benign or serious system failures, in order to observe how the studied OODB behaves and recovers in critical conditions. Such features could be included in VOODB as new modules.

Eventually, to make reusability easier and more formal, VOODB could be rebuilt as part of a reusable model library, as modular fragments that could be assembled to form bigger models. For this sake, slicing the model into fragments is not enough. The structure and interface of each module must also be standardized and an explicit documentation for every sub-model must be provided \cite{BREU98}.

\end{document}